\documentclass[prc,aps,groupedaddress,superscriptaddress,twocolumn,nofootinbib,showpacs,showkeys,floatfix,a4paper,10pt,amsmath,amssymb]{revtex4-1}
\usepackage{graphicx}
\usepackage{dcolumn}
\usepackage{bm}

\begin{document}

\title{Least action description of dynamic pairing correlations in the fission of 
Curium  and Californium isotopes based on the Gogny energy density functional}

\author{R. Rodr\'{\i}guez-Guzm\'an}
\email{raynerrobertorodriguez@gmail.com}
\affiliation{Departamento de F\'isica Aplicada I, Escuela Polit\'ecnica Superior, Universidad de Sevilla, Seville, E-41011, Spain}

\author{L.M. Robledo}
\email{luis.robledo@uam.es}
\affiliation{%
Center for Computational Simulation, Universidad Polit\'ecnica de 
Madrid, Campus Montegancedo, 28660 Boadilla del Monte, Madrid, Spain
}%
\affiliation{Departamento  de F\'{\i}sica Te\'orica and CIAFF, 
Universidad Aut\'onoma de Madrid, 28049-Madrid, Spain}

\author{Carlos A. Jim\'enez-Hoyos}
\email{cjimenezhoyo@wesleyan.edu}
\affiliation{Department of Chemistry, Wesleyan University, Middletown, CT 06459, USA}

\author{N.~C.~Hern\'andez}
\email{norge.us.es}
\affiliation{Departamento de F\'isica Aplicada I, Escuela Polit\'ecnica Superior, Universidad de Sevilla, Seville, E-41011, Spain}

\date{\today}

\begin{abstract} 
The impact of dynamic pairing correlations and their interplay with 
Coulomb antipairing effects on the  systematic of the spontaneous 
fission half-lives for the nuclei $^{240-250}$Cm and $^{240-250}$Cf is 
analyzed, using a hierarchy of approximations based on the 
parametrization D1M of the Gogny energy density functional (EDF). 
First, the constrained Hartree-Fock-Bogoliubov (HFB) approximation is 
used to compute deformed mean-field configurations, zero-point quantum 
corrections and collective inertias either by using the Slater 
approximation to Coulomb exchange and neglecting Coulomb antipairing or 
by fully considering the exchange and pairing channels of the Coulomb 
interaction. Next, the properties of the {\it{least action}} and 
{\it{least energy}} fission paths are compared. In the  {\it{least 
action}} case, pairing is identified as the relevant degree of freedom 
in order to minimize the action entering the Wentzel-Kramers-Brillouin 
(WKB) approximation to the tunneling probability through the fission 
barrier. Irrespective of the treatment of Coulomb exchange and 
antipairing, it is shown that the {\it{least action}} path obtained 
taking into account the pairing degree of freedom leads to stronger 
pairing correlations that significantly reduce the spontaneous fission 
half-lives $t_{SF}$ improving thereby the comparison with the 
experiment by several orders of magnitude. It is also shown that the 
Coulomb antipairing effect is, to a large extent, washed out by the 
{\it{least action}} procedure and therefore the $t_{SF}$ values 
obtained by the two different treatments of the Coulomb exchange 
and pairing are of similar quality.
\end{abstract}


\maketitle{}

%
%
%

\section{Introduction.}

How to account for the most relevant correlations along different 
fission paths of atomic nuclei still represents a major challenge in 
modern nuclear structure physics \cite{ref1,ref1b}. Along its different 
fission paths, the nucleus exhibits pronounced shape changes 
consequence of the subtle balance between Coulomb, surface energy and 
quantum shell effects associated to the underlying single-particle 
structure. Within the mean-field approximation \cite{ref2} often used 
in nuclear physics to characterize nuclear dynamics, those shape 
changes are usually imposed with the help of constrains on  certain 
operators ${\hat{{\bf{Q}}}}$ such as, for example, the quadrupole 
$\hat{Q}_{20}$ and $\hat{Q}_{22}$, axial octupole $\hat{Q}_{30}$, axial 
hexadecupole $\hat{Q}_{40}$ and necking $\hat{Q}_{Neck}$ operators 
\cite{ref1,ref3,ref12}. Microscopic mean-field studies of 
fission-related quantities are usually carried out using both 
non-relativistic 
\cite{ref3,ref4,ref5,ref6,ref7,ref8,ref9,ref10,ref11,ref12,ref13,ref14,ref15,ref16,ref17,ref18,ref19,ref20,ref21} 
and relativistic \cite{ref22,ref23,ref24,ref25,ref26,ref27,ref28} 
energy density functionals (EDFs).

Many of the  mean-field fission studies available in the literature 
make use of the {\it{least energy}} (LE) principle that imposes  each 
{\bf{Q}}-configuration along the fission path  to be determined by 
minimizing the mean-field energy. Short range pairing correlations, of 
great importance for nuclear dynamics, are taken into account within the 
mean-field approach by introducing the Bogoliubov-Valatin canonical 
transformation to quasiparticles that leads to the 
Hartree-Fock-Bogoliubov (HFB) mean-field equations. As a consequence of 
the structure of the quasiparticle operators, particle number 
is not preserved at  the mean-field level and the associated $U(1)$ gauge 
symmetry is broken \cite{ref2}. The amount of pairing correlations 
present in each configuration determines the size of the collective 
inertias associated with shape degrees of freedom and therefore have a 
strong influence in fission observables like spontaneous fission 
have-lives  $t_{SF}$ or fragment mass distributions. The LE  scheme, based on several 
parametrizations of the Gogny-EDF \cite{ref4}, has been employed to 
study the  fission properties of even-even Ra, U, Pu and super-heavy 
nuclei \cite{ref3,ref13,ref29,ref30} as well as odd-mass U, Pu and No 
nuclei \cite{ref31,ref32} within the Equal Filling Approximation (EFA) 
\cite{ref33}. Those previous HFB studies 
\cite{ref3,ref29,ref30,ref31,ref32} revealed that modifications of a 
few percent in the pairing strength of the interaction can have a 
significant impact on the collective masses, leading to uncertainties 
of several orders of magnitude in the predicted  $t_{SF}$ values.

\begin{figure*}
\includegraphics[width=1.00\textwidth]{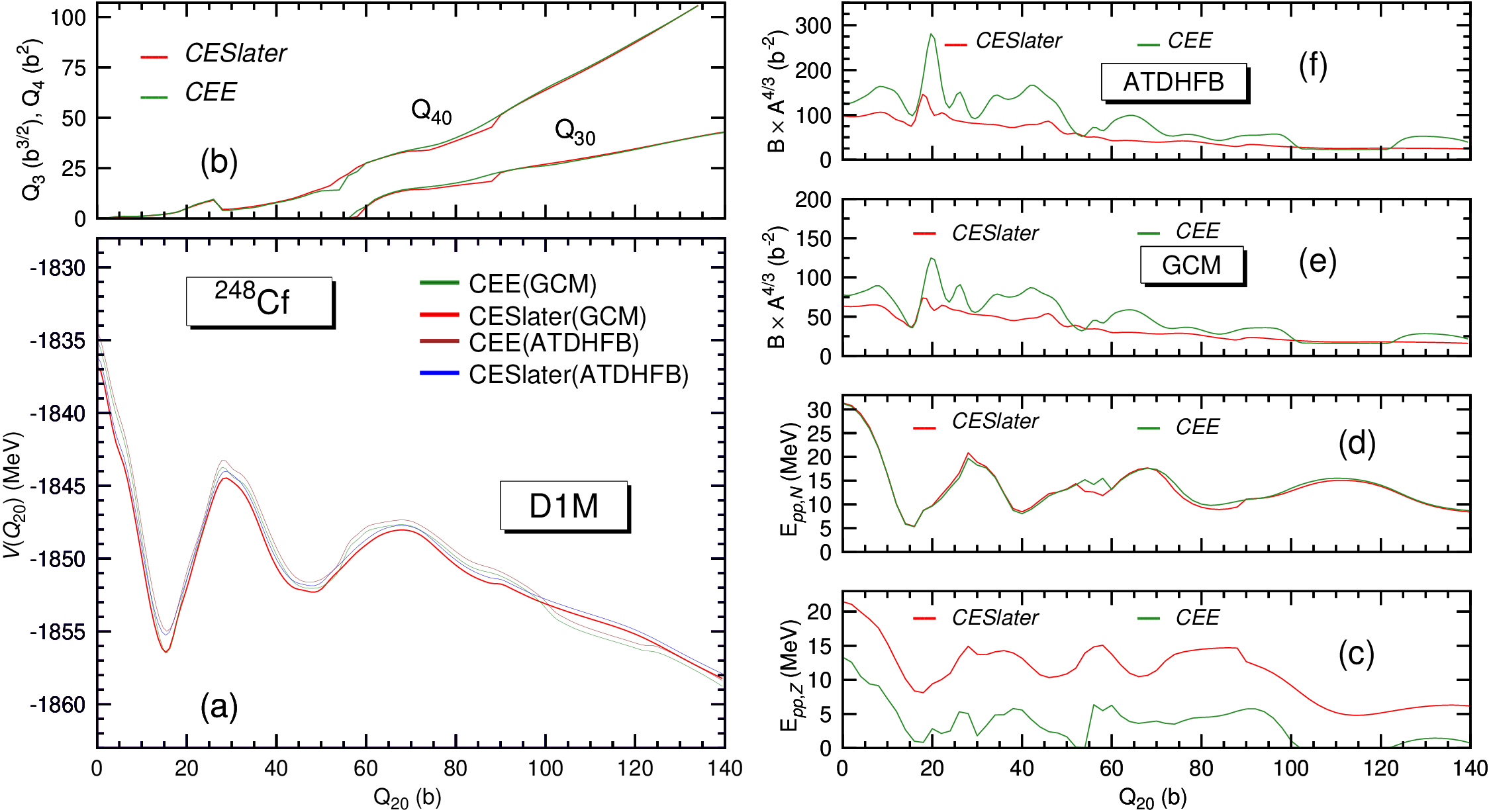}
\caption{(Color online) The  CESlater and CEE {\it{least energy}} (LE) collective potentials
$V(Q_{20})$ Eq.(\ref{coll-potential-V}) 
obtained for the 
nucleus $^{248}$Cf, within the GCM and ATDHFB schemes, are plotted
in panel (a) as functions of the quadrupole moment $Q_{20}$. 
The octupole $Q_{30}$  and hexadecupole $Q_{40}$ 
moments  of the intrinsic states are plotted 
in panel (b). The proton $E_{pp,Z}$ and neutron $E_{pp,N}$ pairing interaction energies 
are depicted in panels (c) and (d), while the
collective 
GCM and ATDHFB masses  are plotted in panels 
(e) and (f). Results have been obtained with the parametrization D1M of the Gogny-EDF. 
}
\label{example248Cf} 
\end{figure*}

The impact of dynamic (beyond-mean-field) pairing correlations on 
fission properties has  been recently studied 
\cite{ref48,rayner-2022-PNP-fission} using a LE restricted variation 
after particle number projection (RVAP-PNP) approach. A  variational 
mean-field subspace has been  built for each {\bf{Q}}-configuration 
along the fission path by using constraints on particle number 
fluctuation for both protons and neutrons separately. The optimal 
mean-field state is then determined by searching for the minimum of the 
particle-number-projected energy in this subspace. The main outcome of 
the first study mentioned before was that the increase of pairing correlations 
associated with RVAP-PNP is compensated by its quenching due to 
Coulomb antipairing and the final results are very similar to those of 
the mean-field. At this point it is worth to remember that Coulomb 
antipairing is neglected in mean-field studies but it has to be fully 
taken into account in the PNP procedure to avoid the Pauli principle violation 
problem of symmetry restoration \cite{ref50bis}. A later study 
\cite{rayner-2022-PNP-fission} revealed that for nuclei with low and 
narrow fission barriers, the compensation between Coulomb antipairing 
and beyond-mean-field pairing  effects might not be perfect leading to 
a modulation of the $t_{SF}$ values, as functions of the neutron 
number, due to PNP. Thus, for such nuclei, the combined  Coulomb antipairing and LE 
RVAP-PNP effects can not be overlooked, neither can they be accounted 
for by the Coulomb Slater approximation \cite{ref61}.

Dynamic pairing correlations, have also been studied within
the {\it{least action}} (LA) principle. 
Here, at variance with the LE scheme, each configuration along the 
fission path is  determined by the minimum of the  action ${\cal{S}}$ 
written in terms of a set of selected collective variables. The motivation
for the use of the LA principle is that the transmission probability through
the fission barrier is proportional to the exponential of the action.
It is well known in the literature \cite{ref34,ref35} that the LE and LA schemes 
provide similar results when only shape constrains are considered. 
However, the situation is rather different when pairing degrees of 
freedom are taken into account in the minimization of the action 
\cite{ref35,ref44}. In this case, the competition between the rapid changes in the 
collective inertia (decreasing as the inverse of the pairing gap 
\cite{ref36,ref37}) and the energy (increasing as the square of the 
pairing gap) leads to a minimum of the action ${\cal{S}}$ for a pairing 
gap far larger than the one corresponding to the minimum-energy 
configuration. The value of the action at the minimum is much smaller 
than its value at the minimum-energy configuration. It is precisely, 
this reduction of the action that leads to a decrease  of the predicted 
LA half-lives as compared to the LE ones 
\cite{ref34,ref35,ref38,ref40,ref41,ref42,ref43,ref44,sugerido-ref-Sta-1,sugerido-ref-Sta-2} bringing them in 
closer agreement with experimental data. Furthermore, the well-known 
dispersion in the predicted $t_{SF}$ values arising from the use of the 
Generator Coordinate Method (GCM) and/or Adiabatic Time Dependent HFB 
(ATDHFB) masses \cite{ref3} is significantly reduced \cite{ref44} 
within the LA scheme. It has also been shown, that LA  pairing fluctuations can restore 
axial symmetry along the fission path \cite{ref42,ref72}. 

\begin{table} 
\label{barriers}
\begin{tabular}{c|cc|cc|cc|cc}
\hline 
         & $B_{I,A}$ & $B_{II,A}$ & $B_{I,B}$ &  $B_{II,B}$ & $B_{I,C}$ & $B_{II,C}$ & $B_{I,D}$ & $B_{II,D}$ \\ \hline 
CESlater &    11.74  &   8.23     & 11.05     &  7.40       & 16.70     &   11.15   &    18.17  &    10.94   \\ \hline 
CEE      &    12.67  &   8.73     & 11.72     &  7.60       & 19.81     &   12.76   &    20.72  &    12.07   \\ \hline
\end{tabular}
\caption{Heights, in MeV, of the inner ($B_{I}$) and outer ($B_{II}$) 
fission barriers  obtained within the {\it{least energy}} (LE) and 
{\it{least action}} (LA) GCM and ATDHFB schemes for $^{248}$Cf. Results 
are shown for the CESlater and CEE approaches with the Gogny-D1M EDF. 
The labels $A$ and $B$ correspond to the GCM and ATDHFB LE paths, whereas labels $C$ and $D$ 
are for the GCM and ATDHFB LA paths, respectively. For details, see the main text.}
\end{table}

The results already mentioned, clearly point towards the need for better 
understanding of the impact of LA pairing correlations in fission calculations, especially 
in those cases where calculations are based on 
sophisticated microscopic Gogny and/or Skyrme EDFs.
Such correlations have already  been considered in previous LA
Gogny-EDF  studies \cite{ref35,ref44}. Nevertheless, in all those 
calculations, the Coulomb exchange term has been considered in the Slater 
approximation \cite{ref61} while Coulomb and spin-orbit antipairing 
are neglected (this approximation will be referred to hereafter as
CESlater).  

In this paper, we extend our previous LA studies \cite{ref35,ref44} to 
a  set of Cm and Cf nuclei for which experimental data are available 
\cite{ref53}. The selected nuclei, i.e., $^{240-250}$Cm and  
$^{240-250}$Cf belong to a region of the nuclear chart where several 
key features of the shell effects associated with super-heavy elements 
start to manifest  \cite{ref13,ref31}. One goal of this work is to 
examine the role of  pairing correlations in the LA framework for such 
nuclei using the CESlater approximation. Our second and most relevant goal is to examine 
the interplay between Coulomb antipairing effects and dynamic pairing 
correlations within the LA scheme. To this end, calculations have been 
carried out including all the direct, exchange and pairing 
contributions coming from the Gogny-EDF \cite{ref4}. In particular, the 
Coulomb exchange and pairing terms are treated exactly (this 
approximation will be referred to hereafter  as CEE). 

The quenching of pairing correlations due to the Coulomb interaction 
(Coulomb antipairing) is usually neglected due to the computational 
effort associated with the evaluation of Coulomb's pairing field. 
However, Coulomb antipairing effects lead to a severe  reduction of the 
pairing gap \cite{ref48,rayner-2022-PNP-fission,ref51,ref52} and 
therefore collective inertias will increase when this contribution is 
included, leading to an increase of the collective action and larger 
$t_{SF}$ values. On the other hand, dynamic pairing correlations tend 
to increase the pairing gap reducing thereby the inertias, leading to a 
decrease of the action and smaller $t_{SF}$ values. The combined effect 
of Coulomb antipairing and dynamic pairing correlations has already 
been considered within the LE RVAP-PNP approach 
\cite{ref48,rayner-2022-PNP-fission}. In the present study, however, we 
consider the interplay between Coulomb antipairing and dynamic pairing 
correlations using a LA perspective. To the best of our knowledge, at 
least in the case of Gogny-like EDFs, the CSlater LA results discussed 
in this paper are the first of their kind reported in the literature 
for $^{240-250}$Cm and  $^{240-250}$Cf. Furthermore, ours is the first  
systematic study of the interplay between Coulomb antipairing and 
dynamic pairing effects within the CEE LA framework.

The results discussed in this paper have been obtained with the 
parametrization D1M  \cite{ref57} of the Gogny-EDF \cite{ref4}. The 
Gogny force has been chosen because its central part is finite range and this guarantees a 
consistent treatment of long range and pairing correlations within the 
same framework. The parametrization D1M has already been shown to 
account reasonably well for fission-related properties in previous 
studies (see, for example, 
Refs.~\cite{ref4,ref29,ref30,ref31,ref32,ref35,ref44,ref57}). This is 
the reason driving its choice as a reference in the present study. 
However, in order to examine the robustness of the results with respect 
to the underlying Gogny-EDF, calculations have also been carried out 
with the D1S \cite{ref4} and D1M$^{*}$ \cite{ref56} parametrizations. 
The predicted D1S and D1M$^{*}$ trends are rather similar to the ones 
obtained with the Gogny-D1M EDF. Therefore, only Gogny-D1M results will 
be discussed in detail in this paper.

The paper is organized as follows. The methodology employed in this 
study is briefly outlined in Sec.~\ref{Theory}. In particular, in this 
section we outline the  LE  and LA schemes  employed to compute the 
paths and spontaneous fission half-lives for the considered nuclei at 
the CESlater and CEE levels. The results of our calculations are 
discussed in Sec.~\ref{results}. First, in Sec.~\ref{pedagogical}, we 
illustrate the employed methodology in the case of $^{248}$Cf. The 
systematic of the fission paths and spontaneous fission half-lives, 
obtained with each of the considered approaches, is discussed in 
Sec.~\ref{systematics-fp} for $^{240-250}$Cm and  $^{240-250}$Cf. 
Finally, Sec.~\ref{conclusions} is devoted to the concluding remarks.


\section{Theoretical framework}
\label{Theory}


In this section, we briefly outline the methodology employed in this 
study \cite{ref3,ref44,rayner-2022-PNP-fission} to obtain LE  and LA 
fission paths as well as other fission-related quantities -- see Refs. 
\cite{ref3,ref44,rayner-2022-PNP-fission} for a more detailed account. 

The starting point in all our considerations is the HFB mean-field 
method implemented for the finite range Gogny force \cite{ref57}. Aside 
from the usual HFB constrains on both the proton ${\hat{Z}}$ and 
neutron ${\hat{N}}$ number operators \cite{ref2}, constrains on the 
mean value of the (axially symmetric) quadrupole $\hat{Q}_{20}$ and 
octupole $\hat{Q}_{30}$   operators have been employed to obtain the 
CESlater and/or CEE  LE (static) paths. The mean values of other shape 
parameters like the hexadecupole $\hat{Q}_{40}$ and higher 
multipolarity operators are given by the Ritz-variational procedure 
\cite{ref2} that determines each of the configurations along the  LE 
paths. Parity is allowed to be broken at any stage of the calculations 
and a constrain on the operator $\hat{Q}_{10}$ is employed to prevent spurious 
effects associated to the center of mass motion \cite{ref3}. 

We are aware of the role of triaxiality around the top of the static 
inner barriers (see, for example, Refs.~\cite{ref3,ref5,ref22}). 
However, it has also been found \cite{ref18,ref26} that the lowering of 
the inner barrier, due to triaxiality, comes together with an increase 
in the collective inertia that tends to compensate in the calculation 
of the action. This result suggests that the axially symmetric path 
would be the preferred one in fission dynamics. Moreover, it has also 
been shown, that pairing fluctuations can restore axial symmetry along 
the fission path \cite{ref42,ref72}. Therefore, the impact of 
triaxiality in the  spontaneous fission half-lives seems to be very 
limited and it has not been considered in this study.

\begin{table} 
\label{tsf-values248Cf}
\begin{tabular}{c|cc|cc}
\hline 
         & $\log_{10} t_\textrm{SF,A}$ & $\log_{10} t_\textrm{SF,B}$ & $\log_{10} t_\textrm{SF,C}$ & $\log_{10} t_\textrm{SF,D}$  \\ \hline 
CESlater &     21.31                   &   25.82                     &    10.57                    & 12.42   \\ \hline 
CEE      &   29.01                     &   36.73                     &    13.90                    & 14.40    \\ \hline
\end{tabular}
\caption{LE and LA  spontaneous fission half-lives (in seconds) 
obtained with the GCM and ATDHFB collective inertias ($E_{0}$ = 0.5 
MeV) for $^{248}$Cf. Results are shown for the CESlater and CEE 
approaches with the Gogny-D1M EDF. The labels A and B correspond to the 
LE case with the GCM and ATDHFB inertias, whereas 
labels C and D 
are for the LA case with the GCM and ATDHFB inertias, respectively. 
For details, see the main text. }
\end{table}

A constrain on the (total) particle number fluctuation  $\Delta 
\hat{N}^2$ operator has also been added to obtain the CESlater and/or 
CEE  LA paths (see, below). In principle, constrains on both the proton 
and neutron  number fluctuation operator should be considered separately
\cite{ref48,rayner-2022-PNP-fission,sugerido-ref-Sta-2}. However, 
as in previous Gogny-like calculations \cite{ref35,ref44} and
in order to alleviate the already substantial computational effort, 
especially at the CEE level, we have restricted to a constrain on the 
total particle number fluctuation operator. Nevertheless, we  have 
also checked that considering a separate constraint on particle number fluctuations for
protons and neutrons does not bring much as compared to the variation of 
the total particle number fluctuation.

The HFB quasiparticle creation and annihilation operators  have been 
expanded in an axially symmetric (deformed) harmonic oscillator  (HO) 
basis containing  states with $J_{z}$ quantum numbers up to 35/2 and up 
to 26 quanta in the z direction. The basis quantum numbers are 
restricted by the condition
\begin{equation}
2 n_{\perp} + |m| + q\, n_{z} \le M_{z, \mathrm{MAX}}
\end{equation}
with $M_{z,\mathrm{MAX}}=17$ and $q=1.5$. This choice is well suited 
for the elongated prolate shapes  typical of the fission process 
\cite{ref3}. In addition, we have optimized the HO length parameters for
each value of the quadrupole moment as to minimize the HFB energy.
An approximate second order gradient method  has been used 
for the solution of the HFB equation \cite{Robledo-Bertsch2OGM} as it 
provides a fast and robust convergence to the HFB solution.

Within the Wentzel-Kramers-Brillouin (WKB) formalism  the LE and LA 
spontaneous fission half-life $t_{SF}$  (in seconds) has been computed  
\cite{ref3,ref12,ref44} following the standard approach
\begin{equation} \label{TSF-WKB}
t_{SF} = 2.86 \times 10^{-21} \times \left(1 + e^{2\cal{S}} \right)
\end{equation}
where the action ${\cal{S}}$ along the (one-dimensional $Q_{20}$-projected) 
fission path is the integral 
\begin{equation} \label{Action}
S = \int_{a}^{b} dQ_{20} {\cal{S}}(Q_{20})
\end{equation}
between the classical turning points $a$ and $b$ of the action 
${\cal{S}}(Q_{20})$ for each quadrupole moment
\begin{equation} \label{Integrand-Action}
{\cal{S}}(Q_{20}) = \sqrt{2 B(Q_{20})\left(V(Q_{20})-\left(E_{Min}+E_{0} \right)  \right)}
\end{equation}

In Eq.(\ref{Action}), the integration limits $a$ and $b$ represent 
classical turning points \cite{proportional-1} for the potential 
$V(Q_{20})$  corresponding to the energy $E_{Min}+E_{0}$. The energy 
$E_{Min}$ corresponds to the absolute minimum of the considered path, 
while $E_{0}$ accounts for the true ground state energy once quadrupole 
fluctuations are taken into account. Although $E_{0}$ could be obtained
from the curvature and collective inertia around the ground state minimum 
we have taken in this work  
the typical value $E_{0}$ = 0.5 MeV for all the isotopes considered. We have checked 
that other values of $E_{0}$ \cite{ref3}, including those estimated in 
terms of the curvature  of the fission path around the ground state  
and the ground state collective quadrupole inertia \cite{ref30}, do not 
alter the qualitative conclusions of the present study. 

From its definition Eq. (\ref{TSF-WKB}), it is clear that the spontaneous 
fission half-life depends on the different approximations considered in 
the calculation as they impact the  potential energy surfaces and 
collective inertias. It is worth to remark that the collective potential $V(Q_{20})$ in 
Eq.(\ref{Integrand-Action}) is given by the HFB(CESlater) and/or HFB(CEE) energies 
corrected by the corresponding quantum zero-point 
vibrational and rotational energies
\begin{small}
\begin{eqnarray}
\label{coll-potential-V}
V(Q_{20}) = E_{HFB}(Q_{20})- \Delta E_{ROT}(Q_{20})- \Delta E_{VIB}(Q_{20})
\end{eqnarray}
\end{small}
The collective mass $B(Q_{20})$ as well as  
the zero-point vibrational energy correction $\Delta E_{VIB}(Q_{20})$ have been  
computed using 
both the ATDHFB and GCM approaches. In all the computations of the 
$t_{SF}$ values, the wiggles in the collective masses have been 
softened by means of a three point filter \cite{ref3}. The rotational 
energy correction $\Delta E_{ROT}(Q_{20})$  has been computed in terms of the Yoccoz moment of 
inertia \cite{ref1,ref2,ref3,ref44,ER-Lectures,NPA-2002}.

In what follows, we outline the methodology employed to obtain the 
(CESlater and CEE) LE  and LA fission paths. First, 
$(Q_{20},Q_{30})$-constrained HFB(CESlater) calculations have been 
carried out for each of the considered nuclei. The two HO length 
parameters  $b_{z}$ and $b_{\perp}$ characterizing the HO basis have 
been optimized so as to minimize the total mean-field energy for each 
configuration along the LE fission path \cite{ref3}. Using the  
HFB(CESlater)  wave functions as starting input, 
$(Q_{20},Q_{30})$-constrained HFB(CEE) calculations have then been 
carried out for $^{240-250}$Cm and  $^{240-250}$Cf without further 
optimizing the  lengths of the HO basis to alleviate the computational 
effort.  Zero-point quantum rotational and vibrational energies have 
been added {\it{a posteriori}} to the HFB(CESlater) and HFB(CEE) 
energies to obtain the  collective potentials $V(Q_{20})$ according to 
Eq.(\ref{coll-potential-V}). As  illustrative examples, the CESlater 
and CEE LE  collective potentials $V(Q_{20})$ obtained for the nucleus 
$^{248}$Cf, within the GCM and ATDHFB schemes, are plotted in panel (a) 
of Fig.~\ref{example248Cf} as functions of the quadrupole moment 
$Q_{20}$. 

\begin{figure}
\includegraphics[width=0.48\textwidth]{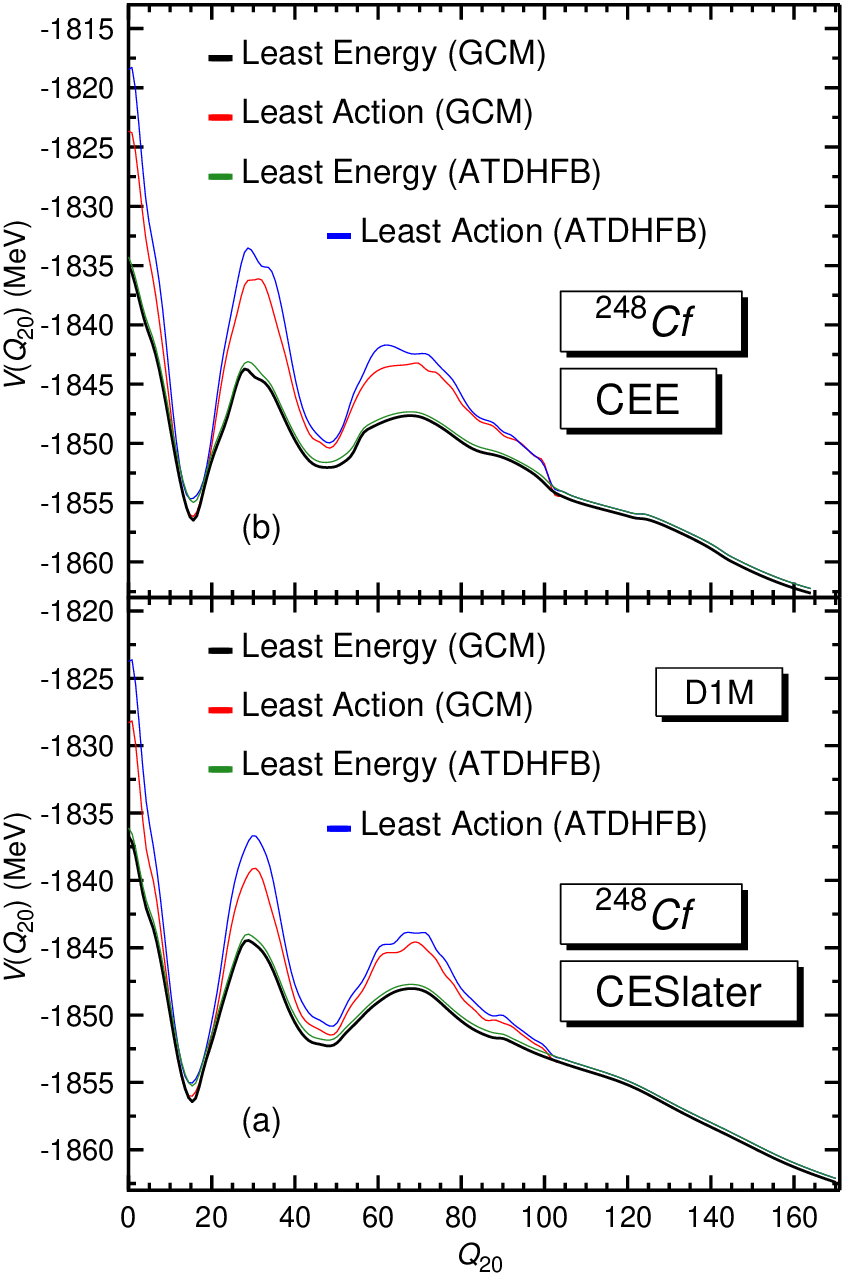}
\caption{(Color online)
The  CESlater and CEE {\it{least action}} (LA) collective potentials
$V(Q_{20})$ Eq.(\ref{coll-potential-V}) 
obtained for the 
nucleus $^{248}$Cf, within the GCM and ATDHFB schemes, are plotted
in panels (a) and (b), as functions of the quadrupole moment $Q_{20}$. 
The  CESlater and CEE {\it{least energy}} (LE) collective potentials
are also included in the plots. Results have been obtained with 
the parametrization D1M of the Gogny-EDF. For more details, see the 
main text.
}
\label{248Cf_ME_MA_CES_CEE_paths} 
\end{figure}


\begin{figure*}
\includegraphics[width=1.0\textwidth]{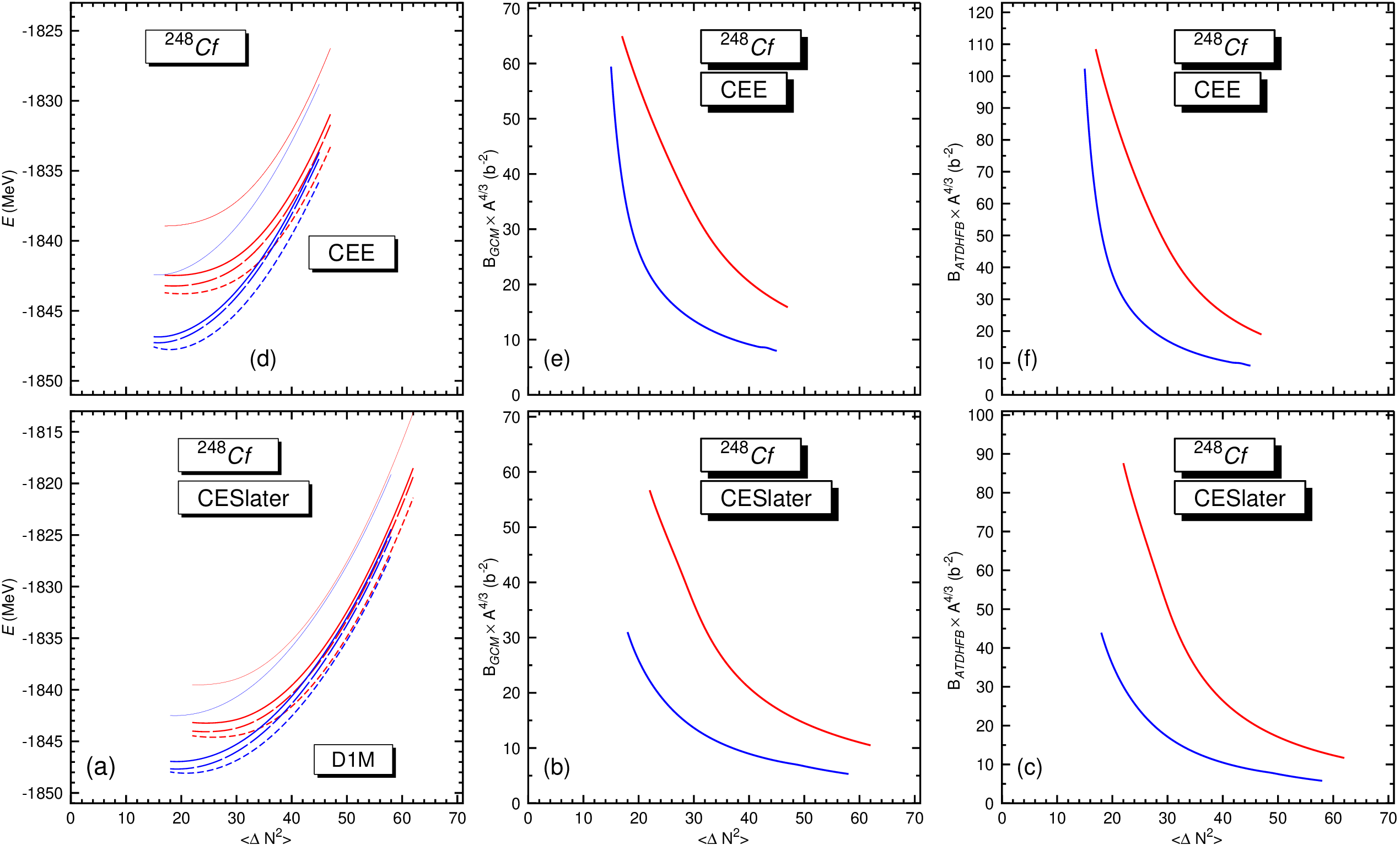}
\caption{(Color online) The intrinsic HFB(CESlater) (full thin lines) 
[HFB(CEE) (full thin lines)], the HFB(CESlater) plus rotational 
correction energies (full thick lines) [HFB(CEE) plus rotational 
correction energies (full thick lines)], the HFB(CESlater) plus 
rotational and vibrational GCM correction energies (short dashed lines) 
[HFB(CEE) plus rotational and vibrational GCM correction energies 
(short dashed lines)] and the HFB(CESlater) plus rotational and 
vibrational ATDHFB correction energies (long dashed lines) [HFB(CEE) 
plus rotational and vibrational ATDHFB correction energies (long dashed 
lines)] are plotted in panel (a) [panel (d)] as functions of $\langle 
\Delta \hat{N}^2 \rangle$. The  CESlater (CEE)  GCM and ATDHFB masses 
are depicted in panels (b) and (c) [(e) and (f) ]. Results are shown 
for the quadrupole moments $Q_{20}$ = 28 (red) and 68 (blue) b. For 
details, see the main text.
}
\label{vs_DN2_248Cf} 
\end{figure*}

The  CESlater and CEE LA paths \cite{ref35,ref44} considered in this 
paper are determined in the following way: for each  
$Q_{20}$-configuration along the CESlater and CEE LE paths,  
$(Q_{20},\Delta {N}^2)$-constrained HFB(CESlater) and  HFB(CEE) 
calculations are performed keeping constant the $Q_{20}$ value and 
equal to the self-consistent solution of the LE path. The range of 
values of the constrained $\langle \Delta \hat{N}^2 \rangle$ quantity 
starts at the self-consistent $\langle \Delta \hat{N}^2 \rangle_{self}$ 
value and extends until a minimum in the corresponding CESlater and CEE 
actions {\it{S}} is reached. We have used a small step size and, once 
more to alleviate the computational effort especially at the CEE level, 
we have optimized as much as possible the number of  $\langle \Delta 
\hat{N}^2 \rangle$ values required to reach a minimum of the action. 
Note, that at the CESlater and CEE level, the minimization is carried 
out for both the actions obtained using GCM and ATDHFB inertias, i.e., 
two  CESlater and CEE dynamic paths are determined \cite{ref44}.

\begin{figure*}
\includegraphics[width=0.90\textwidth]{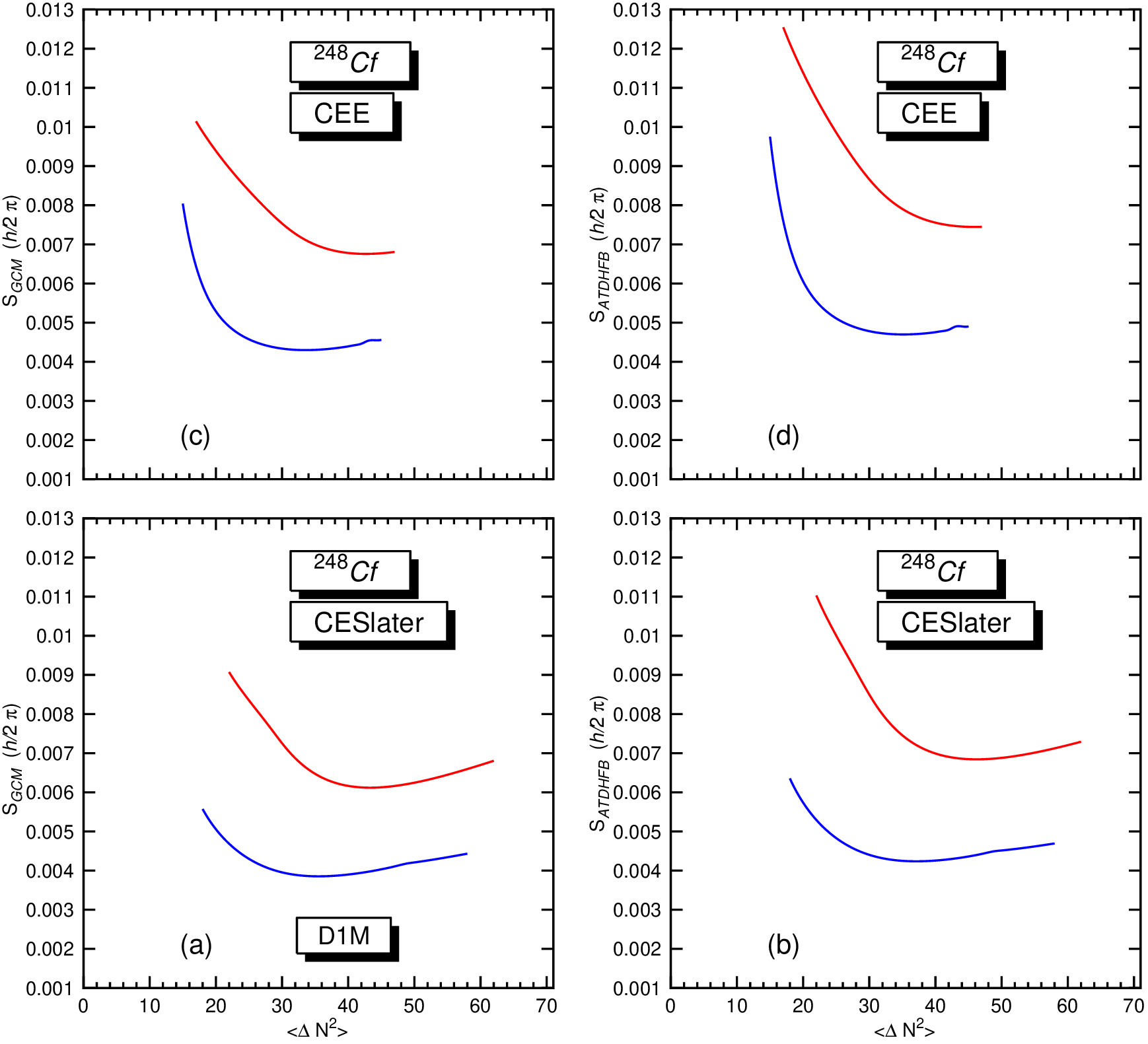}
\caption{(Color online) The GCM and ATDHFB  integrand 
(\ref{Integrand-Action})
of the action (\ref{Action}) corresponding to the CESlater (CEE) approach
are depicted in panels (a) and (b) [(c) and (d)] 
as functions of $\langle \Delta \hat{N}^2 \rangle$. Results are shown 
for the quadrupole moments $Q_{20}$ = 28 (red) and 
68 (blue) b. For details, see the 
main text.
}
\label{Action_DN2_248Cf} 
\end{figure*}


\section{Discussion of the results}
\label{results}

The methodology employed for all the nuclei considered in this work is 
illustrated in the case of $^{248}$Cf and described in detail in 
Sec.~\ref{pedagogical} below. The systematic of the fission paths and 
spontaneous fission half-lives, obtained with each of the considered 
approaches, is discussed in Sec.~\ref{systematics-fp} for 
$^{240-250}$Cm and  $^{240-250}$Cf.

\begin{figure*}
\includegraphics[width=1.0\textwidth]{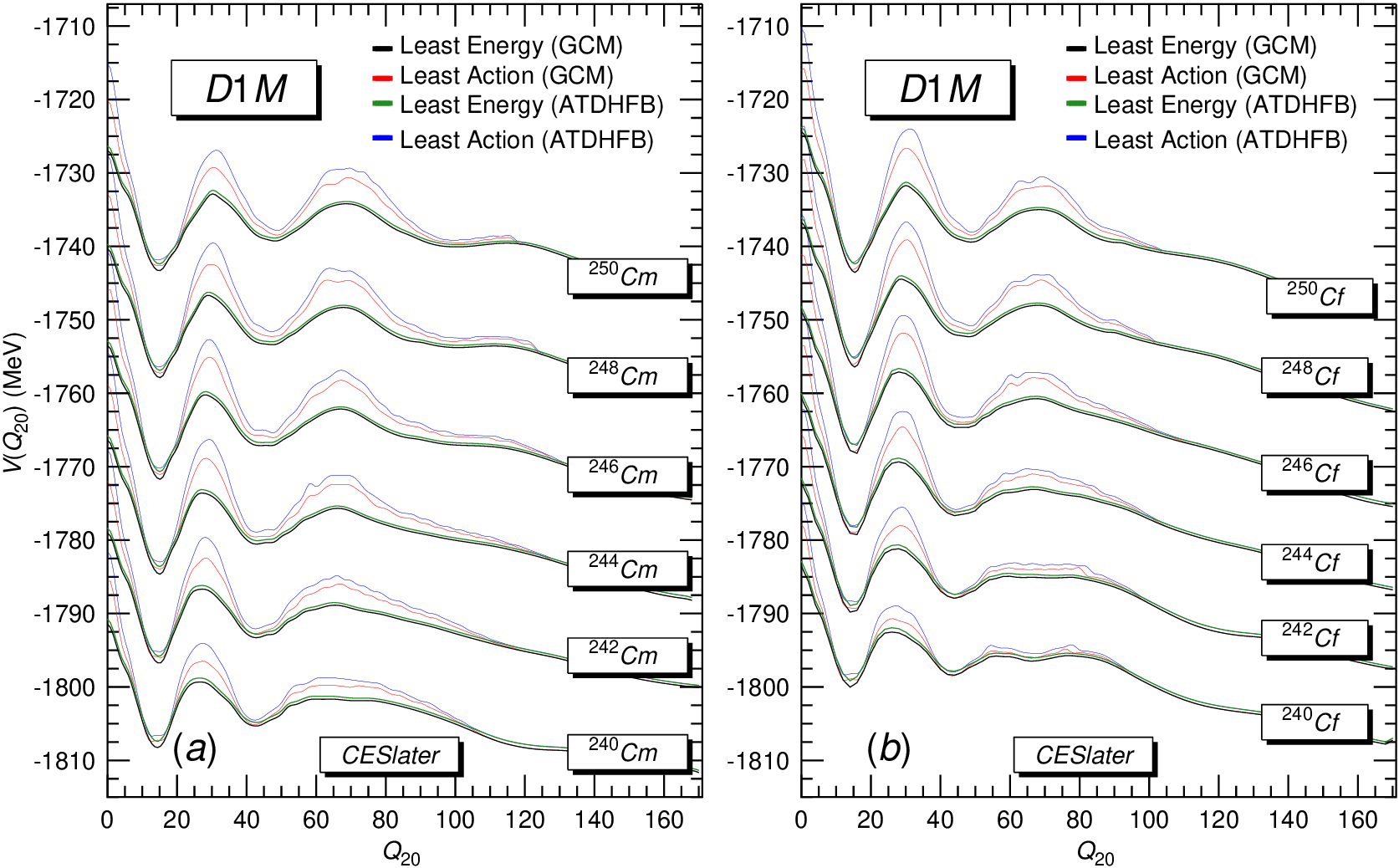}
\caption{(Color online) The  CESlater {\it{least action}} (LA) collective potentials
$V(Q_{20})$ Eq.(\ref{coll-potential-V}) 
obtained for the nuclei $^{240-250}$Cm and 
$^{240-250}$Cf, within the GCM and ATDHFB schemes, are plotted in panels (a) and (b), as functions of the 
quadrupole moment $Q_{20}$. The CESlater {\it{least energy}} (LE) collective potentials
are also included in the plots. The 
paths have been successively shifted by 25 MeV in order to accommodate 
them in a single plot. Results have been obtained with the 
parametrization D1M of the Gogny-EDF. For more details, see the main 
text.
}
\label{FissionPathsCmCf_MA_CSlater} 
\end{figure*}

\subsection{An illustrative example: the nucleus $^{248}$Cf}
\label{pedagogical}

The CESlater and CEE LE  collective potentials $V(Q_{20})$ 
Eq.(\ref{coll-potential-V}) obtained for the nucleus $^{248}$Cf, within 
the GCM and ATDHFB schemes, are plotted in panel (a) of 
Fig.~\ref{example248Cf} as functions of the quadrupole moment $Q_{20}$. 
The absolute minimum of the LE paths is located at $Q_{20} = 16$~b and 
is reflection symmetric $(Q_{30}=0)$. The fission isomer at $Q_{20} = 
48$~b is separated from the ground state by the inner barrier, the top 
of which is  located at $Q_{20} = 26-28$~b. Octupole correlations play 
a prominent  role for quadrupole deformations $Q_{20} \ge 56$~b and 
significantly affect the height of the outer barrier, the top of which 
is located at $Q_{20} = 68$~b. For large quadrupole moments ($Q_{20} 
\ge 100 $~b) the CEE LE collective potentials exhibit a faster decline 
than the CESlater ones \cite{rayner-2022-PNP-fission}. The CESlater and 
CEE LE heights $B_{I,LE}$ and $B_{II,LE}$, computed within both the GCM 
and ATDHFB schemes, are given in Table~\ref{barriers}. The octupole 
$Q_{30}$ and hexadecupole $Q_{40}$ moments of the intrinsic states, 
depicted in panel (b) of Fig.~\ref{example248Cf}, are rather similar in 
the CESlater and CEE approximations.

\begin{figure*}
\includegraphics[width=1.0\textwidth]{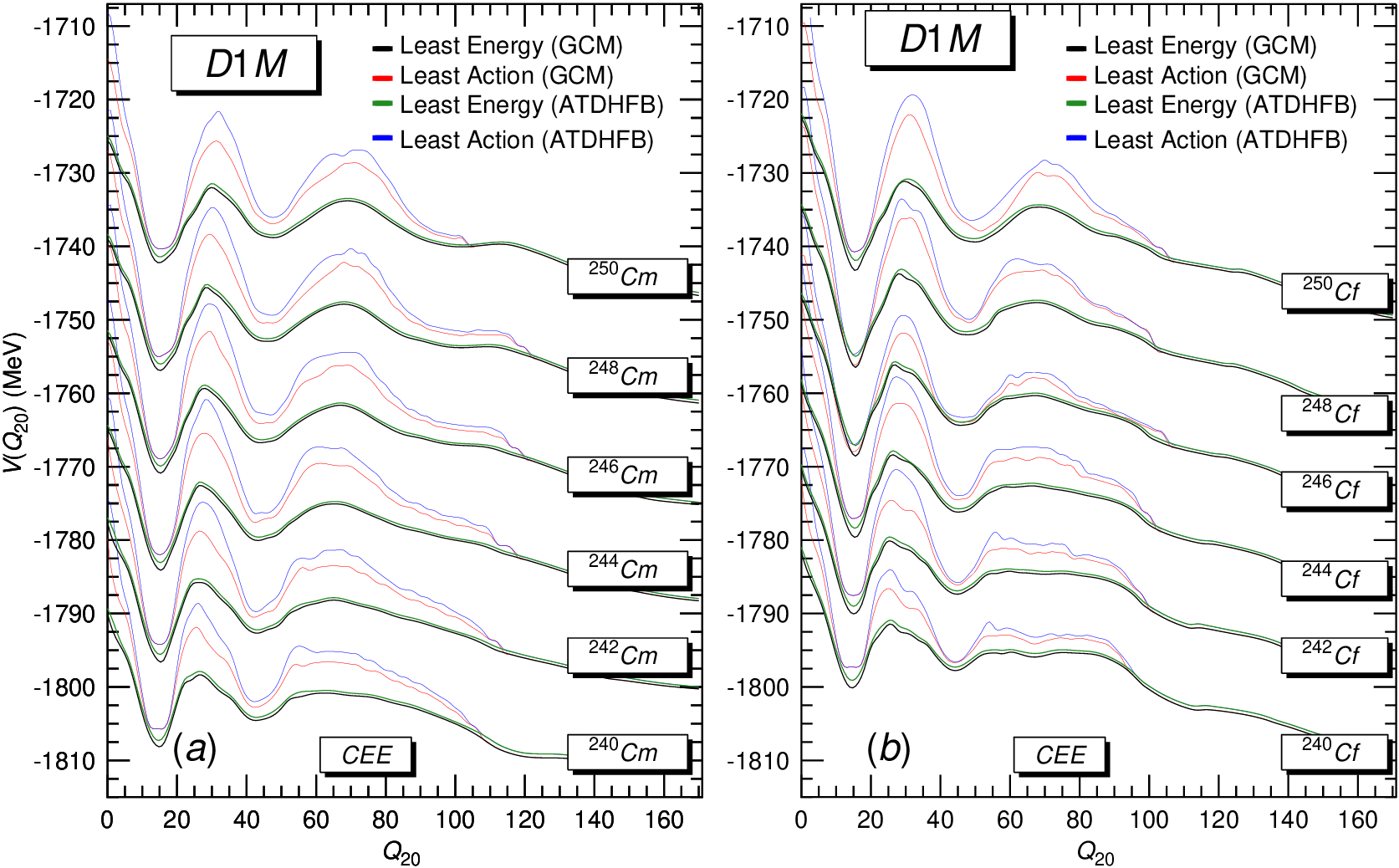}
\caption{(Color online) The  CEE {\it{least action}} (LA) collective 
potentials $V(Q_{20})$ Eq.(\ref{coll-potential-V}) obtained for the 
nuclei $^{240-250}$Cm and $^{240-250}$Cf, within the GCM and ATDHFB 
schemes, are plotted in panels (a) and (b), as functions of the 
quadrupole moment $Q_{20}$. The CEE {\it{least energy}} (LE) collective 
potentials are also included in the plots. The paths have been 
successively shifted by 25 MeV in order to accommodate them in a single 
plot. Results have been obtained with the parametrization D1M of the 
Gogny-EDF. For more details, see the main text.
}
\label{FissionPathsCmCf_MA_CEE} 
\end{figure*}


The HFB(CESlater) and HFB(CEE) proton and neutron  pairing interaction 
energies \cite{ref2} $E_{pp,\tau} = \frac{1}{2} 
\textrm{Tr}(\Delta_{\tau} \kappa_{\tau} )$ (with $\tau=Z$ and $N$) are 
plotted in panels (c) and (d) of Fig.~\ref{example248Cf}. As can be 
seen from panel (c), Coulomb antipairing effect severely quenches the 
proton pairing energies in  the HFB(CEE) states as compared with the  
corresponding HFB(CESlater) values. On the other hand, as can be seen 
from panel (d), the HFB(CESlater) and HFB(CEE) neutron pairing energies 
are, as expected, rather similar.

The collective GCM and ATDHFB masses  are plotted in panels (e) and (f) 
of Fig.~\ref{example248Cf}. Both  masses display a similar trend but 
the ATDHFB masses are, on the average, larger than the GCM ones 
\cite{ref3,ref29,ref30,ref31,ref32,ref44}. This is the reason to 
consider both kinds of collective inertias in this work in the 
calculation of the corresponding LE and LA spontaneous fission 
half-lives. Regardless of the considered GCM and/or ATDHFB scheme, the 
HFB(CEE) collective inertia is larger than the HFB(CESlater) inertia 
and exhibits pronounced high peaks.  This is a consequence of the 
quenching of proton pairing correlations in the HFB(CEE) solutions 
[see, panel (c)] and the inverse dependence of the inertia with the 
square of the pairing gap \cite{ref37}.  


\begin{figure}
\includegraphics[width=0.42\textwidth]{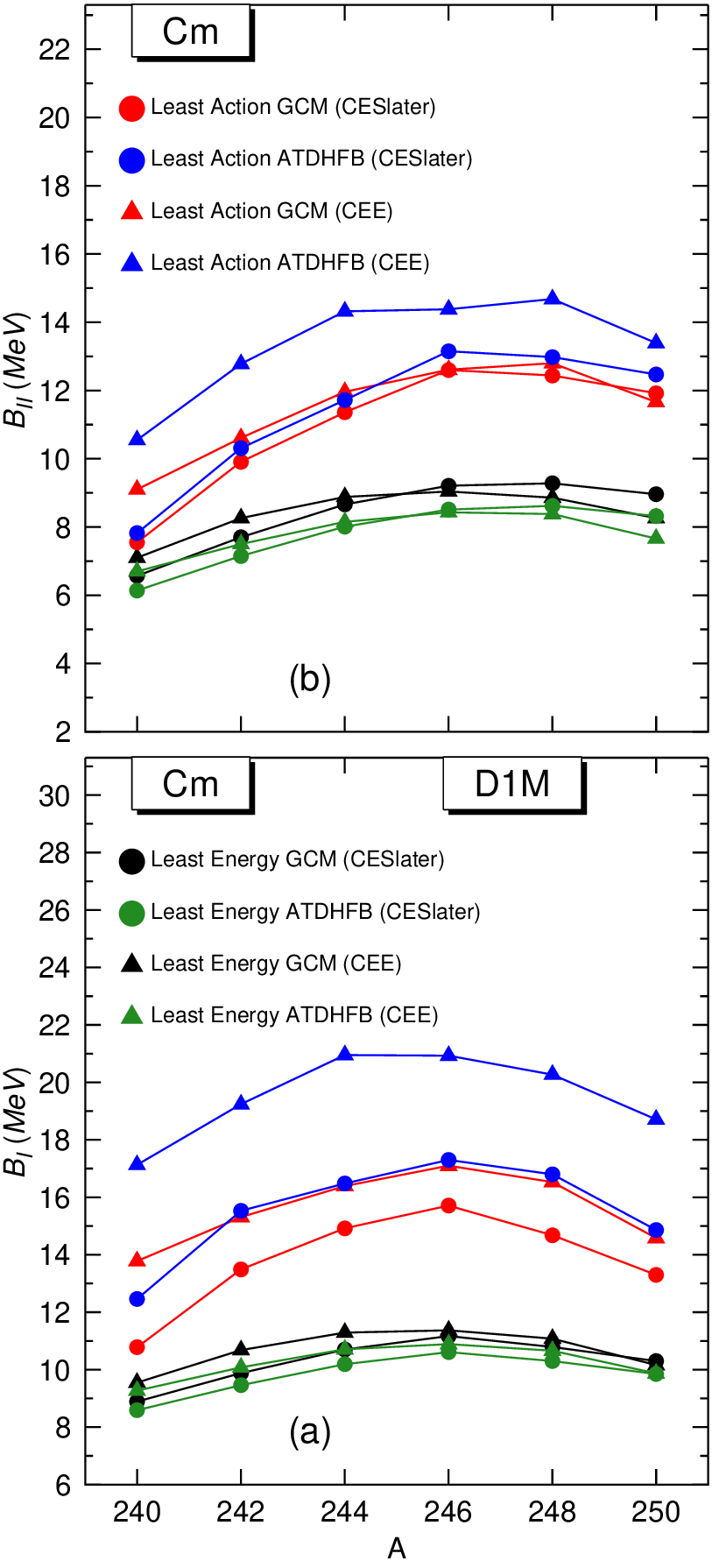}
\caption{(Color online) The CESlater and CEE  heights  $B_{I}$ and 
$B_{II}$ (in MeV) of the inner and outer barriers obtained, within the 
{\it{least energy}} (LE) and {\it{least action}} (LA) GCM and ATDHFB 
schemes, for $^{240-250}$Cm  are plotted in panels (a) and 
(b) as functions of the mass number $A$. Results 
have been obtained with the Gogny-D1M EDF. For details, see the main 
text. 
}
\label{systematic-BI-BII-Cm-Cf} 
\end{figure}


The values of the spontaneous fission half-lives obtained in the LE 
scheme with both  CESlater and  CEE approaches are given in 
Table~\ref{tsf-values248Cf} for  $E_{0}=0.5$ MeV. The  
$t_\textrm{SF,LE}^{ATDHFB}$ values are larger than the 
$t_\textrm{SF,LE}^{GCM}$ ones due to the differences in the 
corresponding collective inertias. Note, that in both the GCM and 
ATDHFB schemes the increase observed in the HFB(CEE) collective 
inertias, due to Coulomb antipairing, leads to a pronounced increase in 
the spontaneous fission half-lives. 

The  CESlater and CEE LA collective potentials $V(Q_{20})$ 
Eq.(\ref{coll-potential-V}) obtained for the nucleus $^{248}$Cf, within 
the GCM and ATDHFB schemes, are plotted in panels (a) and (b) of 
Fig.~\ref{248Cf_ME_MA_CES_CEE_paths}, as functions of the quadrupole 
moment $Q_{20}$. The  CESlater and CEE LE collective potentials are 
also included in the plots for comparison. The deformations of the  
absolute minimum, the top of the inner and outer barriers and the 
fission isomer in the dynamic  paths are similar to the ones obtained 
for the static paths. In the case of the  LA paths, octupole 
correlations  play a key role for $Q_{20} \ge$  54~b. 

As can be seen from panels (a) and (b) of 
Fig.~\ref{248Cf_ME_MA_CES_CEE_paths}, the most pronounced differences 
between the LA and LE paths appear around the spherical configuration 
as well as  around the tops of the inner and outer barriers. This 
agrees with results obtained in previous Gogny LA calculations 
\cite{ref35,ref44}. This is also in line with previous results obtained 
by the Warsaw-Lublin group (see, for example, the top panel of Fig.1 in 
Ref.~\cite{sugerido-ref-Sta-2}) albeit with a simpler microscopic 
model. For example, at the CESlater level the  spherical configuration 
lies 19.52 and 19.00 MeV above the LE GCM and ATDHFB absolute minima, 
whereas it lies 27.52 and 31.09 MeV above the LA GCM and ATDHFB 
absolute minima. On the other hand,  at the CEE level  the  spherical 
configuration lies 21.47 and 20.64 MeV above the LE GCM and ATDHFB 
absolute minima, whereas it lies 32.39 and 36.05 MeV above the LA GCM 
and ATDHFB absolute minima.


\begin{figure*}
\includegraphics[width=0.8\textwidth]{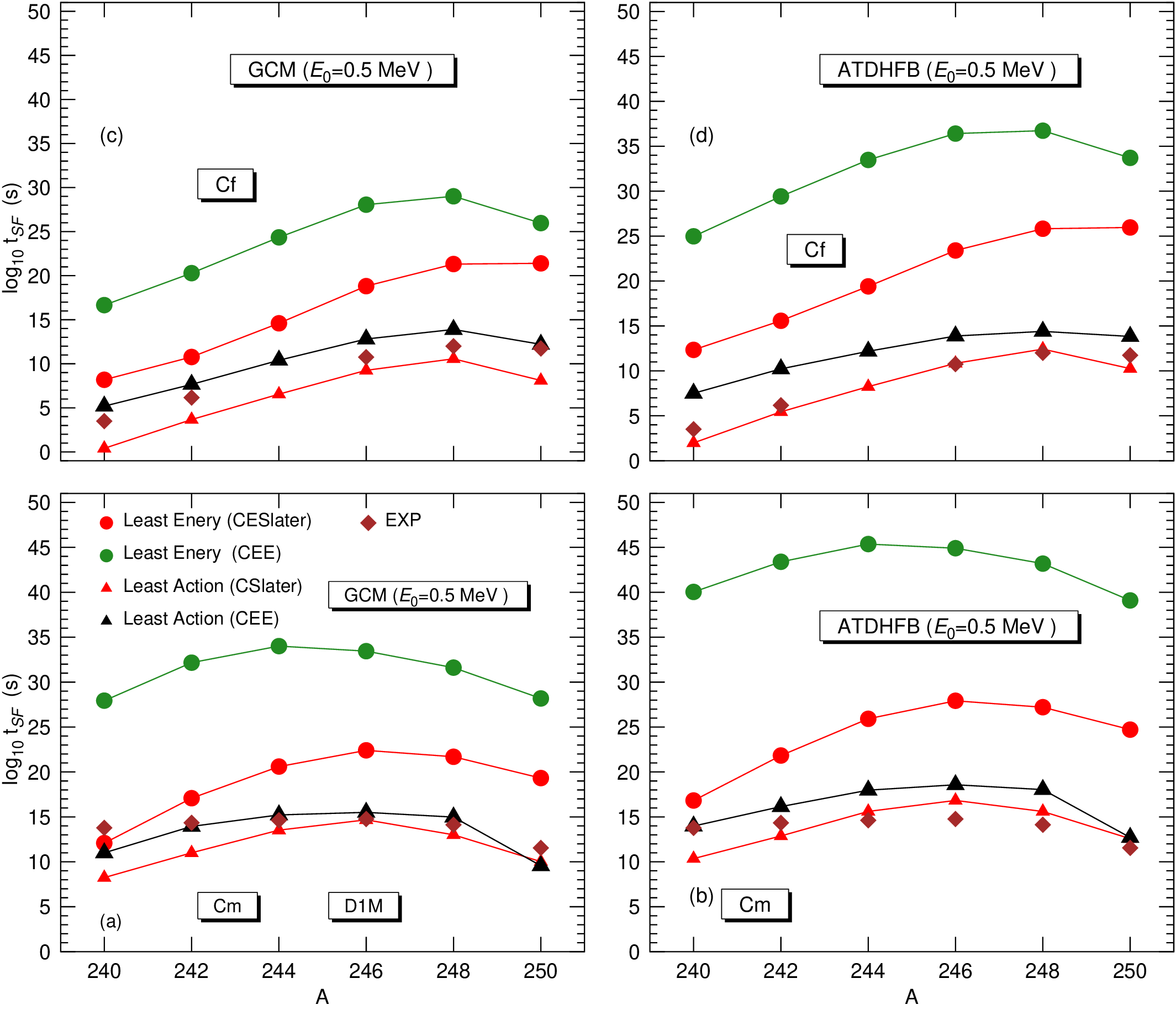}
\caption{(Color online)  The CESlater and CEE spontaneous fission 
half-lives predicted within the {\it{least action}} (LA) GCM and ATDHFB 
schemes for the isotopes $^{240-250}$Cm   are depicted, as functions of 
the mass number $A$, in panels (a) and (b). The  $t_{SF}$ values 
obtained for $^{240-250}$Cf are plotted in panels (c) and (d). 
Calculations have been carried out with E$_{0}$ = 0.5 MeV. Results 
corresponding to the {\it{least energy}} (LE) GCM and ATDHFB schemes  are 
also included in the plot. The experimental $t_{SF}$ values are taken 
from Ref.~\cite{ref53}. 
}
\label{TSF-Cm-Cf_MA} 
\end{figure*}


The  LA GCM and ATDHFB heights  of the inner and outer barriers, 
obtained at the CESlater and CEE levels, are also included in 
Table~\ref{barriers} for $^{248}$Cf. The comparison between the LA and 
LE values given in the table, indicates that the former are always 
larger than the latter. This comparison should be done with care as the 
LE and LA values come from different approximations. On the one hand, 
contrary to cross sections, fission barriers are not experimental 
observables. Fission barriers (in particular, LE fission barriers) are 
usually inferred from cross sections using certain model assumptions. 
One of this assumptions relates with the particular form of the 
collective inertia considered. Within this context, the big differences 
in the table already raise the question about the suitability of a 
direct comparison between  LE  barrier heights and the experimentally 
determined ones (see also, Ref.~\cite{Bertsch2015} and references 
therein) since, the role and behavior of the collective inertia should 
also be taken into account. We will come back to this point later on in 
the paper.

As already mentioned, our goal within the LA scheme is to minimize the action 
Eq.(\ref{Action}). To this end (see, 
Sec.~\ref{Theory}), $(Q_{20},\Delta 
{N}^2)$-constrained  HFB(CESlater) and HFB(CEE) calculations have been 
performed, for each $Q_{20}$-configuration along the LE CESlater and 
CEE fission paths of $^{248}$Cf. In those calculations, we have always 
started at the self-consistent value $\langle \Delta \hat{N}^2 
\rangle_{self}$  for each $Q_{20}$ deformation. However, in order to 
illustrate the inner workings of the LA scheme, it is also
useful to examine the behavior of energies, collective masses and the 
integrand ${\cal{S}}(Q_{20})$ Eq.(\ref{Integrand-Action}) of the action 
${\cal{S}}$ Eq.(\ref{Action}), as functions of $\langle \Delta 
\hat{N}^2 \rangle$ \cite{ref44}. Those quantities are plotted in 
Figs.~\ref{vs_DN2_248Cf} and \ref{Action_DN2_248Cf} for  quadrupole 
deformations $Q_{20}$ = 28 and 68 b corresponding to the top of the 
inner and outer barriers, respectively. 

The intrinsic HFB(CESlater) energies, the HFB(CESlater) plus rotational correction 
energies, the HFB(CESlater) plus rotational and vibrational GCM correction energies
and the HFB(CESlater) plus rotational and vibrational ATDHFB correction energies
are plotted in panel (a) of Fig.~\ref{vs_DN2_248Cf}. The CEE 
energies are shown in panel (d). For each of the considered $Q_{20}$ 
values all those energies, i.e., the intrinsic and the ones including 
zero-point quantum fluctuations, exhibit an almost 
quadratic behavior as functions of $\langle \Delta \hat{N}^2 \rangle$ 
with a minimum at $\langle \Delta \hat{N}^2 \rangle$ = $\langle \Delta 
\hat{N}^2 \rangle_{self}$.

The  CESlater GCM and ATDHFB masses are depicted in panels (b) and (c) 
of Fig.~\ref{vs_DN2_248Cf}, while the CEE masses are depicted in panels 
(e) and (f). Regardless of the  differences in size, the collective 
masses always decrease for increasing $\langle \Delta \hat{N}^2 
\rangle$ values, reflecting their inverse dependence on the square of 
the pairing gap discussed in Refs. \cite{ref36,ref37}.

The CESlater GCM and ATDHFB  integrands ${\cal{S}}(Q_{20})$ of the 
action ${\cal{S}}$ are shown in panels (a) and (b) of 
Fig.~\ref{Action_DN2_248Cf}, while the corresponding CEE quantities are 
plotted in panels (c) and (d), respectively. The panels clearly 
illustrate the main mechanism at play within the LA framework, i.e., the  
competition between the increase of energies 
[panels (a) and (d) of Fig.~\ref{vs_DN2_248Cf}] and the decrease of the 
collective masses [panels (b), (c), (e) and (f) of 
Fig.~\ref{vs_DN2_248Cf}] as functions of $\langle \Delta \hat{N}^2 
\rangle$, leads to a minimum of the CESlater and CEE integrand 
${\cal{S}}(Q_{20})$  for a $\langle \Delta \hat{N}^2 \rangle$ value 
larger than $\langle \Delta \hat{N}^2 \rangle_{self}$. Moreover, as 
expected, the value of the integrand ${\cal{S}}(Q_{20})$ at the minimum 
is smaller than the one at $\langle \Delta \hat{N}^2 \rangle_{self}$. 
As an example, for the configuration  at the top of the inner barrier 
($Q_{20}$ = 28 b) we have $\langle \Delta \hat{N}^2 
\rangle_{self,CESlater}$ = 22 which gives 
${\cal{S}}_{GCM,CESlater}(Q_{20}=28b)$ = 9.07 $\times$ $10^{-3}$ 
$\hbar$ and ${\cal{S}}_{ATD,CESlater}(Q_{20}=28b)$ = 11.03 $\times$ 
$10^{-3}$ $\hbar$. On the other hand, 
${\cal{S}}_{GCM,CESlater}^{min}(Q_{20}=28b)$ 
(${\cal{S}}_{ATD,CESlater}^{min}(Q_{20}=28b)$) reaches a minimum value 
of 6.12 $\times$ $10^{-3}$ $\hbar$ (6.84 $\times$ $10^{-3}$ $\hbar$) at 
$\langle \Delta \hat{N}^2 \rangle$ = 43 (46). 
Similar results are 
obtained at the CEE level. 
Here, for $\langle \Delta \hat{N}^2 
\rangle_{self,CEE}$ = 17, one obtains ${\cal{S}}_{GCM,CEE}(Q_{20}=28b)$ 
= 10.14 $\times$ $10^{-3}$ $\hbar$ and 
${\cal{S}}_{ATD,CEE}(Q_{20}=28b)$ = 12.55 $\times$ $10^{-3}$ $\hbar$. 
The  values at the minima are ${\cal{S}}_{GCM,CEE}^{min}(Q_{20}=28b)$ = 
6.76 $\times$ $10^{-3}$ $\hbar$ and 
${\cal{S}}_{ATD,CEE}^{min}(Q_{20}=28b)$ = 7.45 $\times$ $10^{-3}$ 
$\hbar$ and they correspond to  $\langle \Delta \hat{N}^2 \rangle$ = 43 
and 46.

The quenching of the   GCM and ATDHFB  integrand ${\cal{S}}(Q_{20})$, 
leads to a reduction of the corresponding actions that appear in the 
exponential of the spontaneous fission half-lives Eq.(\ref{TSF-WKB}). 
Thus, the impact on the $t_{SF}$ values is exponential in character. We 
have computed, the CESlater and CEE  LA $t_{SF}$ values  using the GCM 
and ATDHFB inertias. For $E_{0}=0.5$ MeV, the values obtained at the 
CESlater and  CEE levels are given in Table~\ref{tsf-values248Cf}. 
These values should be compared with the corresponding LE lifetimes as 
well as with the experimental value \cite{ref53}  for $^{248}$Cf. From 
the $t_{SF}$ results for $^{248}$Cf given in Table~\ref{tsf-values248Cf}
one can conclude the following 
\begin{itemize}
\item both at the CESlater and CEE levels, the reductions  in the LA 
GCM and ATDHFB spontaneous fission half-lives, bring them closer to the 
experimental value \cite{ref53}. 

\item the differences in the predicted LE $t_{SF}$ values arising from 
the use of GCM and/or ATDHFB masses  are significantly reduced  within 
the LA scheme both at the CESlater and CEE levels. 

\item within the LE scheme, regardless of the employed GCM and/or 
ATDHFB mass, the Coulomb antipairing effect leads to CEE spontaneous 
fission half-lives much larger than the ones obtained at the CESlater 
level. However, such discrepancy is cured, to a large extent, within 
the dynamic description, i.e., the LA scheme provides, regardless of 
the employed collective mass, CESlater and CEE $t_{SF}$ values, which 
are essentially of the same quality.
\end{itemize}

\subsection{Systematic of fission paths and spontaneous fission half-lives
in $^{240-250}$Cm and  $^{240-250}$Cf}
\label{systematics-fp}

The  CESlater LA collective potentials
$V(Q_{20})$ Eq.(\ref{coll-potential-V}) 
obtained for the nuclei $^{240-250}$Cm and 
$^{240-250}$Cf, within the GCM and ATDHFB schemes, are plotted in panels (a) and (b)
of Fig.~\ref{FissionPathsCmCf_MA_CSlater}, as functions of the 
quadrupole moment $Q_{20}$. The corresponding CEE  LA collective potentials
are depicted in panels (a) and (b) of 
Fig.~\ref{FissionPathsCmCf_MA_CEE}. The CESlater and CEE LE collective potentials
are also included in the plots. The paths have been successively shifted by 
25 MeV in order to accommodate them in the plots. We have followed the 
same methodology described in Sec.~\ref{pedagogical} for $^{248}$Cf to 
compute the LA and LE collective potentials shown in the figures.

The reflection symmetric  absolute minima of the CESlater LA and LE 
paths, shown in Fig.~\ref{FissionPathsCmCf_MA_CSlater}, for Cm and Cf 
isotopes correspond to Q$_{20}$ $\approx$ 14-16~b. The fission isomers 
at Q$_{20}$ $\approx$ 42-52~b are separated from the ground state by 
the inner barriers, the top of which are  located at Q$_{20}$ $\approx$ 
24-34~b. Octupole correlations play a significant role for quadrupole 
deformations $Q_{20} \ge 56-62$~b and affect the height of the outer 
barriers, the top of which are located at Q$_{20}$ $\approx$ 54-70~b. 
Similar results hold for the CEE LA and LE paths in 
Fig.~\ref{FissionPathsCmCf_MA_CEE}. Both at the CESlater and CEE 
levels, the most pronounced differences 
between the LE and LA paths are found around the spherical 
configurations as well as around the top of the inner and outer 
barriers.

The CESlater and CEE  heights  $B_{I}$ and 
$B_{II}$  of the inner and outer barriers obtained, within the 
LE and LA GCM and ATDHFB 
schemes, 
for $^{240-250}$Cm  are plotted, in panels (a) and 
(b) of Fig.~\ref{systematic-BI-BII-Cm-Cf} as illustrative examples. For
 the nuclei $^{240-250}$Cf, the barrier 
heights display similar trends 
as functions 
of $A$ and, therefore, they are not shown 
in the plots. As can be seen from the figure, the
CESlater and CEE LA inner and outer barrier 
heights are larger than the LE ones. Larger dynamic 
barrier heights have already been predicted in 
Gogny-CSlater LA calculations \cite{ref35,ref44}. The trends 
observed in Figs.~\ref{FissionPathsCmCf_MA_CSlater}, \ref{FissionPathsCmCf_MA_CEE}
and \ref{systematic-BI-BII-Cm-Cf} also agree well
with 
previous LA calculations (see, for example, Ref.~\cite{sugerido-ref-Sta-2}).
As already discussed in 
Sec.~\ref{pedagogical} [see, panels
(a) and (d) of Fig.~\ref{vs_DN2_248Cf}] , this is a consequence of the fact 
that, for the Q$_{20}$-configurations corresponding to the top of these 
barriers, the energies increase almost quadratically as functions of 
$\langle \Delta \hat{N}^2 \rangle$.

However, caution must be taken in 
connecting the height of fission barriers and spontaneous fission 
half-lives (see, below). First, fission barriers are not physical 
observables but inferred (specially LE barrier heights) 
from cross sections under certain model
assumptions (see \cite{Bertsch2015} and references therein). Second,
for a given nucleus the probability to 
penetrate the fission barrier depends on several other ingredients (for 
example, shape and width of the barrier) and cannot be solely 
determined by the barrier height. Third,
 the behavior and size of the 
collective inertias also play a key role. Obviously,  models 
employing simpler forms of the collective inertias without 
pairing dependences cannot take into account such effects. 
However, as already shown above and 
in previous studies \cite{ref35,ref44,sugerido-ref-Sta-2}, within 
microscopic mean-field
approximations , the increase 
of the  LA barriers, as compared with the LE ones, comes together with 
a dynamic reduction in the corresponding collective inertias and this 
leads to a minimum of the action for those Q$_{20}$-constrained 
configurations. 
 
The CESlater and CEE spontaneous fission half-lives predicted within 
the LA GCM and ATDHFB schemes for the isotopes $^{240-250}$Cm   are 
depicted, as functions of the mass number $A$, in panels (a) and (b) of 
Fig.~\ref{TSF-Cm-Cf_MA}. The  $t_{SF}$ values obtained for 
$^{240-250}$Cf are plotted in panels (c) and (d) of the same figure. 
Calculations have been carried out with E$_{0}$ = 0.5 MeV. Results 
corresponding to the LE GCM and ATDHFB schemes  are also included in 
the plots. The experimental $t_{SF}$ values are taken from 
Ref.~\cite{ref53}. 

On the one hand, both the CESlater and CEE LE approximations already 
account qualitatively  for the experimental trends observed in the 
spontaneous fission half-lives of the nuclei $^{240-250}$Cm and 
$^{240-250}$Cf, as functions of the mass number $A$. Nevertheless, the 
LE approximations overestimate the experimental $t_{SF}$ values 
considerably, especially at the CEE level due to the Coulomb 
antipairing effect. On the other hand, the LA approaches provide, via 
larger dynamic pairing correlations, a severe reduction in the 
predicted  GCM and/or ATDHFB $t_{SF}$ values that brings them closer to 
the experiment \cite{ref53}. For example, in the case of $^{244}$Cm, we 
have obtained  the CESlater (CEE) LE GCM and ATDHFB values $\log_{10} 
t_\textrm{SF,LE}^{GCM}$ = 20.60  (34.01) and $\log_{10} 
t_\textrm{SF,LE}^{ATDHFB}$ = 25.92 (45.36). It is rewarding to notice 
that the corresponding CESlater (CEE) LA GCM and ATDHFB values 
$\log_{10} t_\textrm{SF,LE}^{GCM}$ = 13.53 (15.22) and $\log_{10} 
t_\textrm{SF,LE}^{ATDHFB}$ = 15.61 (17.97) compare much better with the 
experimental spontaneous fission half-life for this nucleus (see, 
Fig.~\ref{TSF-Cm-Cf_MA} and Ref.~\cite{ref53}). 

Moreover, the pronounced differences observed in 
Fig.~\ref{TSF-Cm-Cf_MA} between the LE $t_{SF}$ values computed using 
the GCM and /or ATDHFB inertias \cite{ref3}, with the ATDHFB results 
being larger than the GCM ones, are significantly reduced within the LA 
scheme \cite{ref44}. For example, in the case of $^{244}$Cm, one 
obtains the CESlater (CEE) values $\log_{10} 
\frac{t_\textrm{SF,LE}^{ATDHFB}}{t_\textrm{SF,LE}^{GCM}}$ = 5.32 
(11.35) and $\log_{10} 
\frac{t_\textrm{SF,LA}^{ATDHFB}}{t_\textrm{SF,LA}^{GCM}}$ = 2.08 
(2.75). Note, that the reduction in those differences is even more 
pronounced at the CEE level.

Last but not least, the deficiencies introduced in the CEE LE approach 
by the Coulomb antipairing effect (this effect cannot be properly 
balanced by static pairing correlations \cite{rayner-2022-PNP-fission}) 
are, at least for the studied nuclei, cured to a large extent  within 
the considered LA scheme, where beyond-mean-field pairing correlations 
play a key role when the action is written in terms of pairing degrees 
of freedom. In fact, as can be seen from Fig.~\ref{TSF-Cm-Cf_MA}, 
regardless of the GCM and/or ATDHFB mass employed, the CESlater and CEE 
LA $t_{SF}$ values are essentially of the same quality.


\section{Conclusions}
\label{conclusions}


In this paper, we have examined the role of dynamic pairing 
correlations for a selected set of Cm and Cf isotopes using the 
CESlater LA approximation. Within this context, as in previous studies 
\cite{ref35,ref44}, HFB(CESlater) calculations have been performed 
treating the Coulomb exchange term in the Slater approximation  while 
Coulomb and spin-orbit antipairing have been neglected. The impact of 
Coulomb antipairing and exact Coulomb exchange effects on the LA 
calculations has also been studied. To the best of our knowledge this 
is the first time this type of analysis has been carried out. 
Calculations have  been carried out including all the direct, exchange 
and pairing contributions coming from the Gogny-D1M EDF and the Coulomb 
potential (HFB(CEE)). Constrains on the quadrupole $\hat{Q}_{20}$ and  
the (total) particle number fluctuation  $\Delta \hat{N}^2$ operators 
have  been employed to obtain the LE and LA fission paths. However, it 
should also be kept in mind that octupolarity is allowed to be broken 
at any stage of the calculations. The particle number fluctuation 
$\langle \Delta \hat{N}^2 \rangle$ has been identified as a key degree 
of freedom for the minimization of the WKB action. For the studied 
nuclei it has been shown that, as in previous CESlater calculations 
\cite{ref44}, the  parabolic increase of the energies as functions of  
$\langle \Delta \hat{N}^2 \rangle$ together with the reduction of the 
collective GCM and/or ATDHFB masses lead to a minimum of the  WKB 
action  at a $\langle \Delta \hat{N}^2 \rangle$ value significantly 
larger than the selfconsistent one. This behavior is also present in 
the LA scheme at the CEE level. As a result, both the CESlater and CEE 
LA paths exhibit pronounced differences, as compared with the LE paths, 
around the spherical configurations as well as around the top of the 
inner and outer barriers for all the studied nuclei. Moreover, it has 
been shown that both at the CESlater and CEE levels the LA scheme 
provides, via larger dynamic pairing correlations, a severe reduction 
in the predicted  GCM and/or ATDHFB $t_{SF}$ values that brings them 
closer to the experiment. The pronounced differences between the  LE 
$t_{SF}$ values computed using the GCM and/or ATDHFB inertias are also 
significantly reduced within the LA scheme. Furthermore, it has been 
shown that the strong impact of Coulomb antipairing in the CEE LE 
approach gets reduced to a large extent within the  LA scheme. As a 
consequence, both CESlater and CEE LA $t_{SF}$ values are essentially 
of the same quality.

\begin{acknowledgments}
The work of RR was supported within the framework of the 
(distinguished researcher) Mar\'ia Zambrano Program, Seville University. The  
work of LMR was supported by Spanish Agencia Estatal de Investigacion (AEI)
of the Ministry of Science and Innovation under Grant No. PID2021-127890NB-I00. 
\end{acknowledgments}



\begin{thebibliography}{00}

\bibitem{ref1} N. Schunck and L. M. Robledo, Rep. Prog. Phys.  {\bf{79}}
116301 (2016).

\bibitem{ref1b} H. J. Krappe, and K. Pomorski, Lecture Notes in Physics 838, (Springer Verlag 2012)

\bibitem{ref2}  P. Ring and P. Schuck, 
{\em The Nuclear Many-Body Problem} 
(Springer, Berlin, 1980).

\bibitem{ref3} R. Rodr\'iguez-Guzm\'an and 
L.M. Robledo, Phys. Rev. C {\bf{89}}, 054310 (2014). 

\bibitem{ref12} 
M. Warda and J.L. Egido, 
Phys. Rev. C {\bf{86}}, 014322 (2012).

\bibitem{ref4} 
J. F. Berger, M. Girod, and D. Gogny, 
Nucl. Phys. A {\bf{428}}, 23c (1984).

\bibitem{ref5} 
J.-P. Delaroche, M. Girod, H. Goutte and J. Libert, 
Nucl. Phys. A {\bf{771}}, 103 (2006).

\bibitem{ref6} V. Martin and 
L.M. Robledo, Int. J. Mod. Phys. E {\bf{18}}, 788 (2009).

\bibitem{ref7} 
N. Dubray, H. Goutte and J.-P. Delaroche, 
Phys. Rev. C {\bf{77}}, 014310 (2008).

\bibitem{ref8} S. P\'erez-Mart\'in and 
L.M. Robledo, Int. J. Mod. Phys. E  {\bf{18}}, 861 (2009).

\bibitem{ref9} W. Younes and D. Gogny, Phys. Rev. C {\bf{80}}, 054313 (2009).

\bibitem{ref10} 
M. Warda, J. L. Egido, L.M. Robledo  and K. Pomorski, 
Phys. Rev. C {\bf 66}, 014310 (2002).

\bibitem{ref11} J. L. Egido and L.M. Robledo, Phys. Rev. Lett. {\bf{85}}, 1198
(2000).

\bibitem{ref13} R. Rodr\'iguez-Guzm\'an, Y. M. Humadi 
and L. M. Robledo, Eur. Phys. J. A {\bf{56}}, 43 (2020).

\bibitem{ref14} N. Nikolov, N. Schunck, W. Nazarewicz, M. Bender
 and J. Pei, Phys. Rev. C {\bf{83}}, 034305 (2011).

\bibitem{ref15} 
J.D. McDonnell, W. Nazarewicz and J.A. Sheikh, 
Phys. Rev. C {\bf{87}}, 054327 (2013). 

\bibitem{ref16} 
J. Erler, K. Langanke, H.P. Loens, G. Mart\'inez-Pinedo and P.-G. Reinhard, 
Phys. Rev. C {\bf{85}}, 025802 (2012). 

\bibitem{ref17} 
A. Staszczak, A. Baran, W. Nazarewicz, 
Phys. Rev. C \textbf{87}, 024320 (2013).

\bibitem{ref18} 
A. Baran, K. Pomorski, A. Lukasiak and A. Sobiczewski, 
Nucl. Phys. A {\bf{361}}, 83 (1981).

\bibitem{ref19}  
M. Baldo, L.M. Robledo, P. Schuck and X. Vi\~nas, 
Phys. Rev. C {\bf{87}}, 064305 (2013).

\bibitem{ref20} 
S. A. Giuliani and L.M Robledo, 
Phys. Rev. C \textbf{88}, 054325 (2013).

\bibitem{ref21} Samuel A. Giuliani, Gabriel Martnez-Pinedo, and Luis
M. Robledo Phys. Rev. C {\bf{97}}, 034323 (2018).

\bibitem{ref22} 
H. Abusara, A.V. Afanasjev and P. Ring, 
Phys. Rev. C {\bf{82}}, 044303 (2010).

\bibitem{ref23}  
H. Abusara, A.V. Afanasjev and  P. Ring, 
Phys. Rev. C {\bf{85}}, 024314 (2012).

\bibitem{ref24} 
B.-N. Lu, E.-G. Zhao and S.-G. Zhou, 
Phys. Rev. C {\bf{85}}, 011301 (2012).

\bibitem{ref25} 
S. Karatzikos, A. V. Afanasjev, G. A. Lalazissis and  P. Ring, 
Phys. Lett. B {\bf{689}}, 72 (2010).

\bibitem{ref26} M. Bender, K. Rutz, P.-G. Reinhard, J.A. Maruhn 
and W. Greiner, Phys. Rev. C {\bf{58}}, 2126 (1998). 

\bibitem{ref27} Z. Shi, A. V. Afanasjev, Z. P. Li and J. Meng, Phys. Rev.
C {\bf{99}}, 064316 (2019).

\bibitem{ref28} A. Taninah, S. E. Agbemava and A. V. Afanasjev, Phys.
Rev. C {\bf{102}}, 054330 (2020).

\bibitem{ref29}
R. Rodr\'iguez-Guzm\'an and L. M. Robledo, Eur. Phys. J.
A {\bf{50}}, 142 (2014).

\bibitem{ref30} R. Rodr\'iguez-Guzm\'an and L. M. Robledo, Eur. Phys. J.
A {\bf{52}}, 12 (2016).

\bibitem{ref31} R. Rodr\'iguez-Guzm\'an and L. M. Robledo, Eur. Phys. J.
A {\bf{52}}, 348 (2016).

\bibitem{ref32} R. Rodr\'iguez-Guzm\'an and L. M. Robledo, Eur. Phys. J.
A {\bf{53}}, 245 (2017).

\bibitem{ref33} S. P\'erez-Mart\'in and L.M. Robledo, Phys. Rev. C {\bf{78}},
014304 (2008).

\bibitem{ref48} R. Bernard, S.A. Giuliani and L. M. Robledo, Phys. Rev.
C {\bf{99}}, 064301 (2019).

\bibitem{rayner-2022-PNP-fission} R. Rodr\'iguez-Guzm\'an 
and L. M. Robledo, Phys. Rev. C {\bf{106}}, 024335 (2022).

\bibitem{ref50bis} J A Sheikh, J Dobaczewski, P Ring, L M Robledo and C Yannouleas,
J. Phys. G: Nucl. Part. Phys. 48 123001 (2021).

\bibitem{ref61}
C. Titin-Schnaider and Ph. Quentin, Phys. Lett. B {\bf{49}},
213 (1974).


\bibitem{ref34} J. Sadhukhan, K. Mazurek, A. Baran, J. Dobaczewski,
W. Nazarewicz and J. A. Sheikh, Phys. Rev. C {\bf{88}},
064314 (2013).

\bibitem{ref35} S.A. Giuliani, L. M. Robledo and R. Rodr\'iguez-Guzm\'an,
Phys. Rev. C {\bf{90}}, 054311 (2014).

\bibitem{ref44} R. Rodr\'iguez-Guzm\'an and L. M. Robledo, Phys. Rev. C
{\bf{98}}, 034308 (2018).


\bibitem{ref36} M. Brack, J. Damgaard, A.S. Jensen, H.C. Pauli, V.M
Strutinsky and C.Y. Wong, Rev. Mod. Phys. {\bf{44}}, 320
(1972).

\bibitem{ref37} J.F. Berstch and H. Flocard, Phys. Rev. C {\bf{43}}, 2200
(1991).

\bibitem{ref38} M. Urin and D. Zaretsky, Nucl. Phys. {\bf{75}}, 101 (1976).

\bibitem{ref40} K. Pomorski. Int. J. of Mod. Phys. E {\bf{16}}, 237 (2007).

\bibitem{ref41} A. Staszczak, S. Pilat and K. Pomorski, Nucl. Phys. A
{\bf{504}}, 589 (1989).

\bibitem{ref42} J. Sadhukhan, W. Nazarewicz and N. Schunck, Phys.
Rev. C {\bf{93}}, 011304 (2016).

\bibitem{ref43} J. Zhao, B. -N. Lu, T. Niksic, D. Vretenar and S. -G.
Zhou, Phys. Rev. C {\bf{93}}, 044315 (2016).

\bibitem{sugerido-ref-Sta-1} L. G. Moretto 
and R. P. Babinet, Phys. Lett. B {\bf{49}}, 147 (1974).

\bibitem{sugerido-ref-Sta-2} A. Staszczak, A. Baran, K. Pomorski and 
K. B\"oning, Phys. Lett. B {\bf{161}}, 227 (1985).


\bibitem{ref72} J. Sadhukhan, J. Dobaczewski, W. Nazarewicz, 
J. A. Sheikh and A. Baran,  Phys. Rev. C {\bf{90}}, 061304(R) (2014).

\bibitem{ref53}
N. E. Holden and D. C. Hoffman, Pure Appl. Chem. {\bf{72}},
1525 (2000).

\bibitem{ref51}
T. Lesinski, T. Duguet, K. Bennaceur and J. Meyer, Eur.
Phys. J. A {\bf{40}}, 121 (2009).

\bibitem{ref52} H. Nakada and M. Yamagami, Phys. Rev. C {\bf{83}},
031302(R) (2011).

\bibitem{ref57} 
S. Goriely, S. Hilaire, M. Girod  and S. P\'eru, 
Phys. Rev. Lett. {\bf{102}}, 242501 (2009).

\bibitem{ref56}
C. Gonzalez-Boquera, M. Centelles, X. Vinas, L.M. Robledo,
Phys. Lett. B {\bf{779}}, 195 (2018).

\bibitem{Robledo-Bertsch2OGM} L.M. Robledo and G. F. Bertsch, Phys. Rev. C {\bf{84}}, 014312 (2011).

\bibitem{proportional-1} M. Brack, J. Damgaard, A.S. Jensen, H.C. Pauli, V.M Strutinsky
and C.Y. Wong, Rev. Mod. Phys. {\bf{44}}, 320 (1972).

\bibitem{ER-Lectures} 
J.L. Egido and L.M.Robledo, 
Lectures Notes in Physics {\bf{641}}, 269 (2004).


\bibitem{NPA-2002} 
R. Rodr\'iguez-Guzm\'an, J.L. Egido, and L.M. Robledo,
Nucl. Phys. A {\bf{709}}, 201 (2002).

\bibitem{Bertsch2015} G.F. Bertsch, W. Loveland, W. Nazarewicz,  and P. Talou, 
J. Phys. G Nucl. Part. Phys. {\bf 42}, 077001 (2015)

















\end{thebibliography}
\end{document}